\documentclass[twocolumn,showpacs,preprintnumbers,amsmath,amssymb]{revtex4}
\usepackage{graphicx}
\usepackage{amssymb,amsmath}
\usepackage{bm,slashed}
\usepackage{dcolumn}

\begin{document}

\preprint{}

\title{No Drama Quantum Theory?}

\author{A. Akhmeteli}
\email{akhmeteli@ltasolid.com}
\affiliation{%
LTASolid Inc.\\
10616 Meadowglen Ln. 2708\\
Houston, TX 77042, USA}%


\homepage{http://www.akhmeteli.org}

\date{\today}

\begin{abstract}
This work builds on the following result of a previous article (quant-ph/0509044): the matter field can be naturally eliminated from the equations of the scalar electrodynamics (the Klein-Gordon-Maxwell electrodynamics) in the unitary gauge. The resulting equations describe independent dynamics of the electromagnetic field (they form a closed system of partial differential equations). An improved derivation of this surprising result is offered in the current work. It is also shown that for this system of equations, a generalized Carleman linearization (Carleman embedding) procedure generates a system of linear equations in the Hilbert space, which looks like a second-quantized theory and is equivalent to the original nonlinear system on the set of solutions of the latter. Thus, the relevant local realistic model can be embedded into a quantum field theory. This model is equivalent to a well-established model - the scalar electrodynamics, so it correctly describes a large body of experimental data. Although it does not describe the electronic spin and possibly some other experimental facts, it may be of great interest as a "no drama quantum theory", as simple (in principle) as classical electrodynamics. Possible issues with the Bell theorem are discussed.
\end{abstract}


\maketitle

\section{Introduction}	

Is it possible to offer a "no drama" quantum theory? Something as simple (in principle) as classical electrodynamics - a local realistic theory described by a system of partial differential equations in 3+1 dimensions, but reproducing unitary evolution of quantum theory in the configuration space?

Of course, the Bell inequalities cannot be violated in such a theory. This author has little, if anything, new to say about the Bell theorem, and this article is not about the Bell theorem. However, this issue cannot be "swept under the carpet" and will be discussed in Section IV using other people's arguments.

In a previous article (Ref.~\cite{Akhm10}), the equations of the scalar electrodynamics (the Klein-Gordon-Maxwell electrodynamics) with Lagrangian
\begin{eqnarray}\label{eq:pr6}
\nonumber
-\frac{1}{4}F^{\mu\nu}F_{\mu\nu}+\frac{1}{2}(\psi^*_{,\mu}-ieA_\mu\psi^*)(\psi^{,\mu}+ieA^\mu\psi)-\\
-\frac{1}{2}m^2\psi^*\psi
\end{eqnarray}
were considered:
\begin{equation}\label{eq:pr7}
(\partial^\mu+ieA^\mu)(\partial_\mu+ieA_\mu)\psi+m^2\psi=0,
\end{equation}
\begin{equation}\label{eq:pr8}
\Box A_\mu-A^\nu_{,\nu\mu}=j_\mu,
\end{equation}
\begin{equation}\label{eq:pr9}
j_\mu=ie(\psi^*\psi_{,\mu}-\psi^*_{,\mu}\psi)-2e^2 A_\mu\psi^*\psi.
\end{equation}
Schr\"{o}dinger (Ref.~\cite{Schroed}) noted that the complex charged matter field can be made real by a gauge transform (at least locally), and the equations of motion in the relevant gauge (unitary gauge) for the transformed 4-potential of electromagnetic field $B^{\mu}$ and real matter field $\varphi$ are as follows:
\begin{equation}\label{eq:pr10}
\Box\varphi-(e^2 B^\mu B_\mu-m^2)\varphi=0,
\end{equation}
\begin{equation}\label{eq:pr11}
\Box B_\mu-B^\nu_{,\nu\mu}=j_\mu,
\end{equation}
\begin{equation}\label{eq:pr12}
j_\mu=-2e^2 B_\mu\varphi^2.
\end{equation}
It should be noted that these equations can be obtained from the following Lagrangian (Ref.~\cite{Takabayasi}):
\begin{equation}\label{eq:pr12a}
-\frac{1}{4}F^{\mu\nu}F_{\mu\nu}+\frac{1}{2}e^2 B_\mu B^\mu \phi^2+\frac{1}{2}(\varphi_{,\mu}\varphi^{,\mu}-m^2\varphi^2).
\end{equation}
Actually, it coincides with the Lagrangian of Eq.~(\ref{eq:pr6}) up to the replacement of the complex scalar field by a real one.

The following surprising result was proven in Ref.~\cite{Akhm10}: the equations obtained from Eqs.~(\ref{eq:pr10},\ref{eq:pr11},\ref{eq:pr12}) after natural elimination of the matter field form a closed system of partial differential equations and thus describe independent dynamics of electromagnetic field. The detailed wording is as follows: if components of the 4-potential of the electromagnetic field and their first derivatives with respect to time are known in the entire space at some time point, the values of their second derivatives with respect to time can be calculated for the same time point, so the Cauchy problem can be posed, and integration yields the 4-potential in the entire space-time. Thus, the broad range of quantum phenomena described by the scalar electrodynamics can be described in terms of electromagnetic field only. This result not only permits mathematical simplification, as the number of fields is reduced, but can also be useful for interpretation of quantum theory. For example, in the Bohm (de Broglie-Bohm) interpretation (Refs.~\cite{BohmHiley,Holland,Goldstein}), the electromagnetic field can replace the wave function as the guiding field. This may make the interpretation more attractive, removing, for example, the reason for the following criticism of the Bohm interpretation: "If one believes that the particles are real one must also believe the wavefunction is real because it determines the actual trajectories of the particles. This allows us to have a realist interpretation which solves the measurement problem, but the cost is to believe in a double ontology.~\cite{Smolin}" Independent of the interpretation, quantum phenomena can be described in terms of electromagnetic field only.

The proof in Ref.~\cite{Akhm10} was somewhat awkward. For example, the expression for the matter field contains an ugly looking square root, suggesting sign ambiguities. Therefore, an improved and proof is offered in this work.

It is also shown (using other people's results) how the "one-particle" theories can be turned into "many-particle" theories, which look very much like quantum field theory.

\section{Elimination of Matter Field from Scalar Electrodynamics}

Elimination of the matter field $\varphi$ from Eqs.~(\ref{eq:pr10},\ref{eq:pr11},\ref{eq:pr12}) can be simplified compared to Ref.~\cite{Akhm10}, using a substitution $\Phi=\varphi^2$ first. For example, as
\begin{equation}\label{eq:pr1q}
\Phi_{,\mu}=2\varphi\varphi_{,\mu},
\end{equation}
we obtain
\begin{equation}\label{eq:pr2q}
\Phi_{,\mu}^{,\mu}=2\varphi^{,\mu}\varphi_{,\mu}+2\varphi\varphi^{,\mu}_{,\mu}=
\frac{1}{2}\frac{\Phi^{,\mu}\Phi_{,\mu}}{\Phi}+2\varphi\varphi^{,\mu}_{,\mu}.
\end{equation}
Multiplying Eq.~(\ref{eq:pr10}) by $2\varphi$, we obtain the following equations in terms of $\Phi$ instead of Eqs.~(\ref{eq:pr10},\ref{eq:pr11},\ref{eq:pr12}):
\begin{equation}\label{eq:pr3q}
\Box\Phi-\frac{1}{2}\frac{\Phi^{,\mu}\Phi_{,\mu}}{\Phi}-2(e^2 B^\mu B_\mu-m^2)\Phi=0,
\end{equation}
\begin{equation}\label{eq:pr4q}
\Box B_\mu-B^\nu_{,\nu\mu}=-2e^2 B_\mu\Phi,
\end{equation}
To prove that these equations describe independent evolution of the electromagnetic field $B^\mu$, it is sufficient to prove that if components $B^\mu$ of the potential and their first derivatives with respect to $x^0$ ($\dot{B}^\mu$) are known in the entire space at some time point $x^0=\rm{const}$  (that means that all spatial derivatives of these values are also known in the entire space at that time point), Eqs.~(\ref{eq:pr3q},\ref{eq:pr4q}) yield the values of their second derivatives, $\ddot{B}^\mu$, for the same value of $x^0$. Indeed, $\Phi$ can be eliminated using Eq.~(\ref{eq:pr4q}) for $\mu=0$, as this equation does not contain $\ddot{B}^\mu$ for this value of $\mu$:
\begin{equation}\label{eq:pr5q}
\Phi=(-2e^2 B_0)^{-1}(\Box B_0-B^\nu_{,\nu 0})=(-2e^2 B_0)^{-1}(B^{,i}_{0,i}-B^i_{,i 0})
\end{equation}
(Greek indices in the Einstein sum convention run from $0$ to $3$, and Latin indices run from $1$ to $3$).
Then $\ddot{B}^i$ ($i=1,2,3$) can be determined by substitution of Eq.~(\ref{eq:pr5q}) into Eq.~(\ref{eq:pr4q}) for $\mu=1,2,3$:
\begin{equation}\label{eq:pr6q}
\ddot{B}_i=-B^{,j}_{i,j}+B^\nu_{,\nu i}+(B_0)^{-1} B_i(B^{,j}_{0,j}-B^j_{,j 0}).
\end{equation}
Thus, to complete the proof, we only need to find $\ddot{B}^0$. Conservation of current implies
\begin{equation}\label{eq:pr7q}
0=(B^\mu \Phi)_{,\mu}=B^\mu_{,\mu}\Phi+B^\mu\Phi_{,\mu}.
\end{equation}
This equation determines $\dot{\Phi}$, as spatial derivatives of $\Phi$ can be found from Eq.~(\ref{eq:pr5q}). Differentiation of this equation with respect to $x^0$ yields
\begin{eqnarray}\label{eq:pr8q}
\nonumber
0=(\ddot{B}^0+\dot{B}^i_{,i})\Phi+(\dot{B}^0+B^i_{,i})\dot{\Phi}+\\
+\dot{B}^0\dot{\Phi}+\dot{B}^i\Phi_{,i}+B^0\ddot{\Phi}+B^i\dot{\Phi}_{,i}.
\end{eqnarray}
After substitution of $\Phi$ from Eq.~(\ref{eq:pr5q}), $\dot{\Phi}$ from Eq.~(\ref{eq:pr7q}), and $\ddot{\Phi}$ from  Eq.~(\ref{eq:pr3q}) into  Eq.~(\ref{eq:pr8q}), the latter equation determines $\ddot{B^0}$ as a function of $B^\mu$, $\dot{B}^\mu$ and their spatial derivatives (again, spatial derivatives of $\Phi$ and $\dot{\Phi}$ can be found from the expressions for $\Phi$ and $\dot{\Phi}$ as functions of $B^\mu$ and $\dot{B}^\mu$). Thus, if $B^\mu$ and $\dot{B}^\mu$ are known in the entire space at a certain value of $x^0$, then $\ddot{B}^\mu$ can be calculated for the same $x^0$, so integration yields $B^\mu$ in the entire space-time. Therefore, we do have independent dynamics of electromagnetic field.

Apparently, it is possible to introduce a Lorentz-invariant Lagrangian with higher derivatives that does not include the matter field, but is largely equivalent to the Lagrangian of Eq.~(\ref{eq:pr12a}) (the significance of some special cases, e.g., $\varphi=0$ and $B^\mu B_\mu=0$ (see below) is unclear (different particles?)). To this end, the latter Lagrangian can be expressed in terms of $\Phi=\varphi^2$, rather than $\varphi$, using, e.g., the following:
\begin{equation}\label{eq:pr16a}
\varphi_{,\mu}\varphi^{,\mu}=\frac{1}{4}\frac{\Phi_{,\mu}\Phi^{,\mu}}{\Phi},
\end{equation}
and then $\Phi$ can be replaced by the following expression obtained from the equations of motion Eq.~(\ref{eq:pr4q}):
\begin{equation}\label{eq:pr16b}
\Phi=-\frac{1}{2e^2} \frac{B^\mu(\Box B_\mu-B^\nu_{,\nu\mu})}{B^\mu B_\mu}.
\end{equation}
\section{Transition to Many-Particle Theory}
The theory we considered in the previous section is not second-quantized, so, on the face of it, it cannot describe many particles. On the other hand, nightlight (Ref.~\cite{nightlight2}) indicated that, rather amazingly, second-quantized theories (or at least theories that look like second-quantized ones) can be obtained from nonlinear partial differential equations by a generalization of the Carleman linearization (Carleman embedding) procedure (Ref.~\cite{Kowalski}). This generalized procedure generates for a system of nonlinear partial differential equations a system of linear equations in the Hilbert space, which looks like a second-quantized theory and is equivalent to the original nonlinear system on the set of solutions of the latter.

Following Ref.~\cite{Kowalski2}, let us consider a nonlinear differential equation in an (s+1)-dimensional space-time (the equations describing independent dynamics of electromagnetic field for the scalar electrodynamics are a special case of this equation) ${\partial_t}\boldsymbol{\xi}(x,t) = \boldsymbol{F}(\boldsymbol{\xi},{D^\alpha}\boldsymbol{\xi};x,t)$  , $\boldsymbol{\xi}(x,0)=\boldsymbol{\xi}_0(x)$, where $\boldsymbol{\xi}:\mathbf{R}^s\times\mathbf{R}\rightarrow\mathbf{C}^k$, $D^\alpha\boldsymbol{\xi}=\left(D^{\alpha_1}\xi_1,\ldots ,D^{\alpha_k}\xi_k\right)$, $\alpha_i$ are multiindices, ${D^\beta}={\partial^{|\beta|}}/\partial x_1^{\beta_1}\ldots\partial x_s^{\beta_s}$, with $ |\beta|=\sum\limits_{i=1}^{s}\beta_i$, is a generalized derivative, $\boldsymbol{F}$ is analytic in $\boldsymbol{\xi}$, $D^\alpha\boldsymbol{\xi}$. It is also assumed that $\boldsymbol{\xi_0}$ and $\boldsymbol{\xi}$ are square integrable. Then Bose operators $\boldsymbol{a^\dag(x)}=\left(a^\dag_1(x),\ldots,a^\dag_k(x)\right)$ and $\boldsymbol{a(x)}=\left(a_1(x),\ldots,a_k(x)\right)$ are introduced with the canonical commutation relations:
\begin{eqnarray}\label{eq:ad1}
\nonumber
\left[a_i(x),a^\dag_j(x')\right]=\delta_{ij}\delta(x-x')I,\\
\left[a_i(x),a_j(x')\right]=\left[a^\dag_i(x),a^\dag_j(x')\right]=0,
\end{eqnarray}
where $x,x'\in\mathbf{R}^s$, $i,j=1,\ldots,k$. Normalized functional coherent states in the Fock space are defined as $|\boldsymbol{\xi}\rangle =\exp\left(-\frac{1}{2}\int d^sx|\boldsymbol{\xi}|^2\right)\exp\left(\int d^sx\boldsymbol{\xi}(x)\cdot\boldsymbol{a}^\dagger(x)\right)|\boldsymbol{0}\rangle$. They have the following property: \begin{equation}\label{eq:ad1a}
\boldsymbol{a}(x)|\boldsymbol{\xi}\rangle =\boldsymbol{\xi}(x)|\boldsymbol{\xi}\rangle,
\end{equation}.
Then the following vectors in the Fock space can be  introduced:
\begin{eqnarray}\label{eq:ad2}
\nonumber
|\xi,t\rangle = \exp\left[\frac{1}{2}\left(\int {d^s}x|\boldsymbol{\xi}|^2-\int {d^s}x|\boldsymbol{\xi}_0|^2\right)\right]|\boldsymbol{\xi}\rangle\\
\nonumber
=\exp\left(-\frac{1}{2}\int d^sx|\boldsymbol{\xi}_0|^2\right)\\
\times\exp\left(\int d^sx\boldsymbol{\xi}(x)\cdot\boldsymbol{a}^\dagger(x)\right)|\boldsymbol{0}\rangle.
\end{eqnarray}
Differentiation of Eq.~(\ref{eq:ad2}) with respect to time $t$ yields, together with Eq.~(\ref{eq:ad1a}), a linear Schr\"{o}dinger-like evolution equation in the Fock space:
\begin{eqnarray}\label{eq:ad3}
\nonumber
\frac{d}{dt}|\xi,t\rangle = M(t)|\xi,t\rangle,\\
|\xi,0\rangle=|\boldsymbol{\xi}_0\rangle,
\end{eqnarray}
where the boson "Hamiltonian" $M(t) = \int {d^s}x{\boldsymbol{a}^\dagger}(x)\cdot F(\boldsymbol{a}(x),{D^\alpha}\boldsymbol{a}(x))$.

Can similar results be obtained for the Dirac-Maxwell electrodynamics, based on the results of Ref.~\cite{Akhm10}? Probably yes, but this has not been done yet, and it may require the use of the fermionic coherent states (Ref.~\cite{Cahill}). It should also be emphasized that the results of Ref.~\cite{Akhm10} are less general for the Dirac-Maxwell electrodynamics than for the scalar electrodynamics.

\section{Bell Theorem}
In Section III, it was shown that a theory similar to quantum field theory (QFT) can be built that is basically equivalent to non-second-quantized scalar electrodynamics on the set of solutions of the latter. However, the local realistic theory violates the Bell inequalities, so this issue is discussed below using other people's arguments. Most of them were outlined by nightlight in various forums (see, e.g., Ref.~\cite{nightlight}) and by E. Santos (see, e.g., Ref.~\cite{Santos}), and can be summarized as follows.

While the Bell inequalities cannot be violated in local realistic theories, there are some reasons to believe these inequalities cannot be violated either in experiments or in quantum theory. Indeed, there seems to be  a consensus among experts that "a conclusive experiment falsifying in an absolutely uncontroversial way local realism is still missing"~\cite{Gen}. For example, A. Shimony offers the following opinion:

"The incompatibility of Local Realistic Theories with Quantum Mechanics permits adjudication by experiments, some of which are described here. Most of the dozens of experiments performed so far have favored Quantum Mechanics, but not decisively because of the "detection loophole" or the "communication loophole." The latter has been nearly decisively blocked by a recent experiment and there is a good prospect for blocking the former.~\cite{Shimony}"

M. Aspelmeyer and A. Zeilinger agree:

"But the ultimate test of Bell's theorem is still missing:
a single experiment that closes all the loopholes at once.
It is very unlikely that such an experiment will disagree
with the prediction of quantum mechanics, since this
would imply that nature makes use of both the detection
loophole in the Innsbruck experiment and of the
locality loophole in the NIST experiment. Nevertheless,
nature could be vicious, and such an experiment is desirable
if we are to finally close the book on local realism."~\cite{Aspel}

The popular argument of the latter quote that the loopholes were closed in separate experiments does not look conclusive either. Otherwise one could argue, for example, that the sum of the angles of a triangle in planar Euclidian geometry can differ from 180 degrees because experiments demonstrate that the sum of angles can differ from 180 degrees for planar quadrangles and for triangles on a sphere. The Bell inequalities for local realistic theories can only be guaranteed if all conditions of the Bell theorem are fulfilled simultaneously. Therefore, if one of the assumptions of the Bell theorem is not fulfilled in an experiment, the violation of the Bell inequalities  in that experiment cannot rule out local realistic theories.

On the other hand, to prove theoretically that the inequalities can be violated in quantum theory, one needs to use the projection postulate (loosely speaking, the postulate states that if some value of an observable is measured, the resulting state is an eigenstate of the relevant operator with the relevant eigenvalue). However, such postulate, strictly speaking, is in contradiction with the standard unitary evolution of the larger quantum system that includes the measured system and the measurement device (and the observer, if needed), as such postulate introduces irreversibility, whereas there is no irreversibility for the larger system (see, e.g., Ref.~\cite{alla} or the references to journal articles there), and, according to the quantum recurrence theorem, the larger system will return to a state that can be arbitrarily close to its initial, pre-measurement state, at least in a very large, but finite volume (Ref.~\cite{Bocc}).  Furthermore, unitary evolution cannot generate a mixture of states (the well-known measurement problem in quantum theory). The standard argument that collapse takes place during measurements and unitary evolution takes place between measurements does not seem convincing, as there is no obvious reason why unitary evolution should not be applicable to an instrument or an observer. For example, based on an analysis of experimental data, Schlosshauer (Ref.~\cite{Schloss}) believes that
 "(i) the universal validity of unitary dynamics and the superposition principle has been confirmed far into the mesoscopic and macroscopic realm in all experiments conducted thus far;
(ii) all observed ''restrictions'' can be correctly and completely accounted for by taking into account environmental decoherence effects;
(iii) no positive experimental evidence exists for physical state-vector
collapse;
(iv) the perception of single ''outcomes'' is likely to be explainable through decoherence effects in the neuronal apparatus."

 Therefore, it seems that mutually contradictory assumptions (e.g., unitary evolution and the projection postulate) are required to prove the Bell theorem, so it is on shaky grounds both theoretically and experimentally. On the other hand, the local realistic theory of this work reproduces unitary evolution of a theory that looks like a quantum field theory, so it may need a modification of the theory of measurement (cf. Ref.~\cite{Santos2}).

\section{Conclusion}

Natural elimination of the matter field from the non-second-quantized scalar electrodynamics yields a theory that is, on the one hand, as simple (in principle) as classic electrodynamics, on the other hand, can be embedded into a quantum field theory.

\section*{Acknowledgments}

The author is grateful to A.E. Allahverdyan, A.V. Gavrilin, A.Yu. Kamenshchik, A.Yu. Khrennikov, nightlight, H. Nikoli\textrm{$\mathrm{\acute{c}}$}, W. Struyve, and H. Westman for their interest in this work and valuable remarks.



\vspace*{-5pt}   

\bibliographystyle{unsrt}

\bibliography{akhijqi}

\end{document}